\documentclass[english,twocolumn,superscriptaddress,aps,prb]{revtex4-2}
\usepackage[T1]{fontenc}
\usepackage[utf8]{inputenc}
\usepackage{graphicx}
\usepackage{amssymb}
\usepackage{amsmath}
\usepackage{siunitx} 

\def \bk{{\bf k}}

	\newcommand{\D}{\mathrm{d}}

	\newcommand{\bbm}{\begin{pmatrix}}
	\newcommand{\ebm}{\end{pmatrix}}
	\newcommand{\bma}{\begin{matrix}}
	\newcommand{\ema}{\end{matrix}}
	\newcommand{\bsm}{\begin{smallmatrix}}
	\newcommand{\esm}{\end{smallmatrix}}
	\newcommand{\bsbm}{\left( \begin{smallmatrix}}
	\newcommand{\esbm}{\end{smallmatrix}\right)}
	\newcommand{\To}{\rightarrow}

	\newcommand{\vb}[1]{\left( #1 \right)}
	\newcommand{\vsb}[1]{\left[ #1 \right]}

	\newcommand{\mbf}[1]{\mathbf{#1}}
	\newcommand{\abs}[1]{\vert #1 \vert}

	\newcommand{\dt}{\,\text{.}}

	\newcommand{\usi}[2]{\ensuremath{#1 \, \si{#2}}}

	\newcommand{\com}{\quad\text{,}}

	\usepackage[normalem]{ulem} 
	\usepackage{xcolor}

	\definecolor{mgreen}{RGB}{21,105,26}
	
\usepackage{babel}

\begin{document}

\title{Frequency dependence of the light-induced Hall effect in dissipative graphene}

\author{M. Nuske}
\affiliation{Zentrum f\"ur Optische Quantentechnologien, Universit\"at Hamburg, 22761 Hamburg, Germany}
\affiliation{Institut f\"ur Laserphysik, Universit\"at Hamburg, 22761 Hamburg, Germany}
\affiliation{The Hamburg Center for Ultrafast Imaging, Luruper Chaussee 149, Hamburg 22761, Germany}
\affiliation{Deutsches Forschungszentrum für Künstliche Intelligenz, GmbH (DFKI), Saarbrücken, Germany}
\author{L. Mathey}
\affiliation{Zentrum f\"ur Optische Quantentechnologien, Universit\"at Hamburg, 22761 Hamburg, Germany}
\affiliation{Institut f\"ur Laserphysik, Universit\"at Hamburg, 22761 Hamburg, Germany}
\affiliation{The Hamburg Center for Ultrafast Imaging, Luruper Chaussee 149, Hamburg 22761, Germany}

\begin{abstract}
	We determine the Hall conductivity of light-driven graphene, with  specific focus on its frequency dependence, and compare it to the static effective approximation, based on Floquet states. This approximation gives the Haldane model as the effective model for light-driven graphene, with a gapped spectrum and a quantized Hall conductivity of $-2e^2/h$. We simulate both the light-driven and the effective model, and explicitly include the dissipative environment in our simulations. We investigate the effect of different driving regimes and dissipation strengths on the Hall conductivity in graphene. As a central result, the Hall conductivity of the light-driven system is not well approximated by the effective model, except for a regime of intermediate driving frequencies and small dissipation where the Hall conductivity contribution of the Dirac point approximately recovers the quantized value of $-2e^2/h$, as well as in the transient dynamics for weak dissipation.
\end{abstract}


	
	

\maketitle

\section{Introduction} 
	While topologically insulating states of solids in equilibrium have been established, characterized by a quantized Hall current, a new route of designing materials with a Hall conductivity via optical control is developed in Refs.~\cite{calvo_tuning_2011,usaj_irradiated_2014,perez-piskunow_floquet_2014,kristinsson_control_2016} , and has been demonstrated experimentally in graphene in Ref.~\cite{mciver_light-induced_2019}. Given that optical driving is a form of periodic driving, Floquet theory provides a framework of description for ligth-driven materials, insofar as it provides a natural set of quasi-energy states. Floquet theory therefore provides a mapping of a periodic time-dependent Hamiltonian onto an effective static one \cite{bukov_universal_2015,eckardt_high-frequency_2015}. By using such an effective Hamiltonian Refs.~\cite{oka_photovoltaic_2009,kitagawa_topological_2010,kitagawa_transport_2011,dehghani_out--equilibrium_2015} have proposed that circularly polarized light drives graphene into a topologically insulating state. In the topological regime a gap opens at the Dirac point and the effective Hamiltonian resembles the Haldane model \cite{haldane_model_1988}. Under the additional assumption that the electron distribution forms a band insulating state the system obtains a quantized Hall conductivity of $-2e^2/h$ \cite{oka_photovoltaic_2009,kitagawa_topological_2010}. However, we emphasize that this hypothesis about the light-driven system at high frequencies requires these simplifying assumptions. As we discuss in this paper, the conductivity of the light-driven system is in general not described by this hypothesis, except for narrow parameter regimes. 

	We note that the experimental realization of light-induced Floquet physics is limited to frequencies that are typically smaller than the band widths of the electronic bands. In the case of light-driven graphene, the size of the band gap at the Dirac point scales as $E_{\rm dr}^2/\omega_{\rm dr}^3$, where $E_{\rm dr}$ is the electric field strength and $\omega_{\rm dr}$ the driving frequency. In order to obtain a sizable gap for driving frequencies larger than the broadening of the Floquet or the Bloch states it is necessary to simultaneously increase the field strength to experimentally unfeasible regimes. Therefore the experimental realization of the light-induced Hall effect in graphene has been performed for low driving frequencies of \usi{48}{\tera\hertz} \cite{mciver_light-induced_2019}. We note that, apart from being experimentally more feasible, low-frequency driving also adds to the diversity of the topological phase diagram as additional chern numbers occur at each of the resonances \cite{lindner_floquet_2011,sentef_theory_2015,mikami_brillouin-wigner_2016}.

	To capture the light-induced dynamics and the steady state of the electrons in a solid, it is imperative to include dissipation in its theoretical description. The specific dissipative processes and time scales are to be adjusted to describe a specific material or sample. We note that we have demonstrated that the resulting Hall conductivity measured in Ref.~\cite{mciver_light-induced_2019} can be recovered quantitatively in this manner, see Ref.~\cite{nuske_graphene}. As we point out, the resulting magnitude of the Hall conductivity is not consistent with the assumption that a band insulator state on the Floquet states is formed, which would result in a quantized conductivity that is related to the Chern numbers of the occupied states. Instead, the electron distribution of the steady state is a broad, smooth distribution over the Floquet states, and the Hall conductivity is composed of the weighted group velocity and the weighted Berry curvature of the Floquet states. As such, we refer to it as a geometric-dissipative effect. Other studies were reported in Refs.~\cite{seetharam_controlled_2015,eckardt_atomic_2016,strater_interband_2016,kohler_driven_2005,kohler_floquet-markovian_1997,langemeyer_energy_2014,russomanno_floquet_2017,scopa_lindblad-floquet_2018,schnell_is_2020,dehghani_dissipative_2014,iadecola_occupation_2015,shirai_floquetgibbs_2018,sato_microscopic_2019,sato_floquet_2019,sato_light-induced_2019,broers_observing_2021,broers_detecting_2021,esin_quantized_2018}.

	In this paper we present a more extensive discussion of the effects of dissipation on the light-induced Hall effect in graphene with a particular focus on its behavior with increasing driving frequencies. In addition to the frequency dependence, we present the dependence of the Hall conductivity on the driving field strength and dissipation strength. We find that the Hall conductivity contribution of the Dirac point of the light-driven system is greatly suppressed by dissipation even in the high-frequency limit, which implies that the limit of the effective static model is not achieved. We observe this discrepancy for most of the parameter regimes that we consider, except for small dissipation and intermediate driving frequencies. We observe the same tendency in the transient regime as well, rather than the steady state behavior.

	This paper is organized as follows: in section \ref{numericalFormalism} we describe the formalism employed for all calculations. In sections \ref{highFrequencyLimit}, \ref{lowDissipation} and \ref{transientResponse} we provide a comparison of the light-induced Hall conductivity and the corresponding value for the corresponding static effective model. In section \ref{highFrequencyLimit} we consider the limit of increasing driving frequencies, in section \ref{lowDissipation} the limit of decreasing dissipation and in section \ref{transientResponse} the transient regime for short times. In section \ref{summary} we conclude.

\section{Numerical formalism}\label{numericalFormalism}
	\begin{figure*}[tb]
		\centering
		\includegraphics[width=399pt]{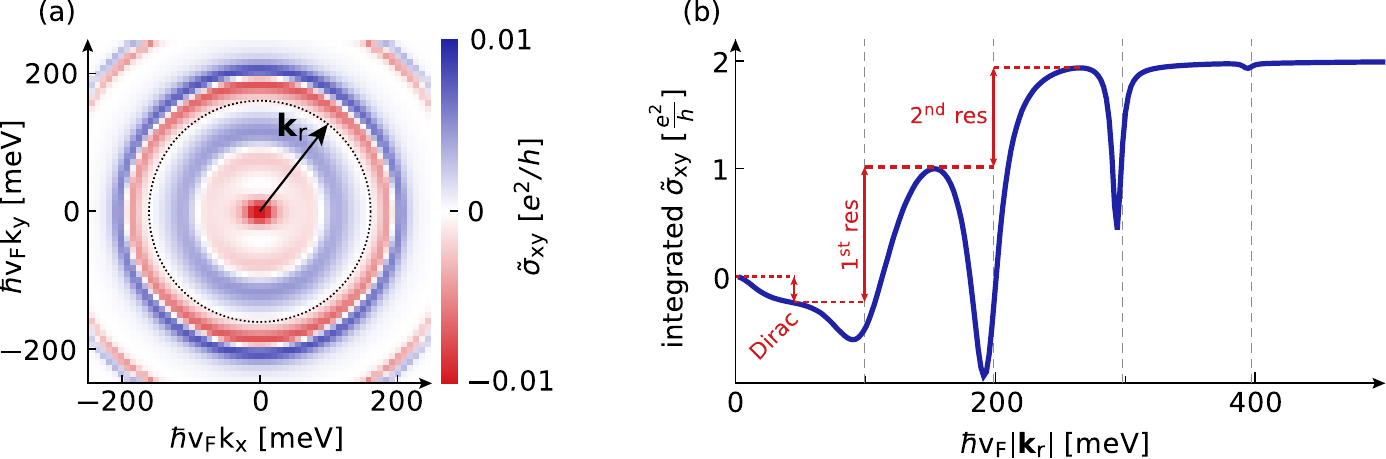}%
		\caption{Circular dichroism of the Hall conductivity. (a) We show the conductivity dichroism $\tilde \sigma_{\rm xy}(\mbf k)$. Note that $\hbar v_Fk=\usi{200}{\milli\eV}$ corresponds to $k\approx\usi{0.03}{\per\angstrom}$.
		Panel (b) shows the same data as (a) but radially integrated as a function of threshold momentum $\abs{\mbf k_{\rm r}}$. The first four resonances are indicated by dashed lines. We oberve that the conductivity contributions are localized around individual resonances and indicate their net contributions by red arrows. For both panels we use $E_{\rm dr}=\usi{20}{\mega\volt\per\meter}$ $\omega_{\rm dr}=\usi{2\pi\cdot 48}{\tera\hertz}\approx\usi{200}{\milli\eV}/\hbar$, $T_1=\usi{1}{\pico\second}$, $T_2=\usi{200}{\femto\second}$, $T_p=\usi{250}{\femto\second}$, $T=\usi{1}{\kelvin}$, $E_{\rm L}=\usi{1.7}{\kilo\volt\per\meter}$, $\mu=0$, $t_{r,\rm dr}=\usi{1}{\pico\second}$ and $\sigma_{r,\rm dr}=\usi{0.5}{\pico\second}$. We show the conductivity density after a steady state is achieved for a tanh-type ramp of the driving field strength.}
		\label{fig:momentumResolvedConductivity}
	\end{figure*}
	Our primary objective is to describe the light-induced dynamics of electrons in graphene. Our discussion takes the experimental regime of Ref.~\cite{mciver_light-induced_2019} as its starting point, and strongly expands its parameter range to determine if and under what conditions the hypothesized high-frequency limit is achieved, in which the Floquet quasi-energy states are interpreted as the energy states of a static effective Hamiltonian. Here we describe the key features of our formalism, for a detailed description see Ref.~\cite{nuske_graphene}.

	To determine the electron dynamics of graphene we integrate the time evolution of the density matrix $\rho$. We factorize the density matrix $\rho$ as $\rho = \prod_\bk \rho_\bk$, on a discrete lattice of $N\times N$ momenta $\bk$, centered around the Dirac point. At each momentum $\bk$ we represent $\rho_\bk$ in the four-dimensional basis of $\Psi_\bk$ consisting of $|0\rangle$, $c_{\bk,A}^\dagger |0\rangle$, $c_{\bk,B}^\dagger |0\rangle$, $c_{\bk,A}^\dagger c_{\bk,B}^\dagger |0\rangle$. Here the operators $c_{\bk,C}^\dag$ and $c_{\bk,C}$ create and annihilate an electron on the $C=A,B$ sublattice, respectively. We solve the master equation for each $\rho_\bk$. In the master equation we include unitary contributions based on the Hamiltonian 
	\begin{align*}
		H = H_{0,k} + H_{dr,k}(t) + H_{L,k}(t)\com 
	\end{align*}
	in which the contributing terms are given in Eqs.~\ref{eq:undrivenHamiltonian}, \ref{eq:drivingHamiltonian} and \ref{eq:dcHamiltonian}. The low-energy dynamics of electrons in graphene is described by Dirac Hamiltonians of the form 
	\begin{align}
		H_{0,\mbf k}&= \Psi_k^\dag \Big[ \hbar v_{\rm F}\vb{ \tau_z k_x \sigma_x + k_y \sigma_y} \Big] \Psi_k \label{eq:undrivenHamiltonian}\com
	\end{align}
	where $v_{\rm F}=\usi{10^6}{\meter\per\second}$ is the Fermi velocity, $\sigma_i$, $i=x,y,z$, denote Pauli matrices with respect to the singly-occupied sector and $\tau_z=\pm 1$ is the valley index.
	Additionally we include the light-matter interaction $H_{\rm em,\bk}$ which consists of a circularly-polarized driving term
	\begin{align}
		H_{\rm dr,\mbf k}(t)&= \frac{ev_{\rm F}E_{\rm dr}}{\omega_{\rm dr}} \nonumber\\
		&\Psi_k^\dag \vsb{ \tau_z \sin\vb{\omega_{\rm dr}t} \sigma_x - \sigma_{\rm pol} \cos(\omega_{\rm dr}t) \sigma_y } \Psi_k \label{eq:drivingHamiltonian}
	\end{align}
	and a longitudinal DC probing field
	\begin{align}
	 	H_{\rm L, \mbf k}&= - \tau_z e v_{\rm F} E_{\rm L}t\; \Psi_k^\dag \sigma_x \Psi_k \label{eq:dcHamiltonian} \dt
	\end{align} 
	For later convenience we give the Floquet theory result for the effective static Hamiltonian for Eq.~\ref{eq:drivingHamiltonian}, at second order in the driving field $E_{\rm dr}$. In the high-frequency limit \cite{oka_photovoltaic_2009,kitagawa_topological_2010}
	\begin{align}
		H_{\rm eff,\mbf k}&=- {\sigma_{\rm pol} \Delta_{\rm hf}} \; \Psi_k^\dag \sigma_z \Psi_k \com \label{eq:effHighFreqHamiltonian}
	\end{align}
	where $\Delta_{\rm hf}=\vb{\hbar v_{\rm F} e E_{\rm dr}}^2/{(\hbar \omega_{\rm dr})^{3}}$. For our simulations we include an additional switch-on function of the form 
	\begin{align}
		\frac{\tanh\vb{\frac{t-t_{r,\alpha}}{\sigma_{r,\alpha}}}+1}2\com
	\end{align}
	with the time scales $t_{r,\alpha}$ and $\sigma_{r,\alpha}$ as a global prefactor of the driving, the longitudinal probing term and the effective Hamiltonian. We use different switch-on time scales for the longitudinal term ($\alpha=\rm L$) and the driving and effective Hamiltonian ($\alpha=\rm dr$). For the longitudinal probing term $t_{r,L}=\usi{50}{\femto\second}$ and $\sigma_{r,L}=\usi{20}{\femto\second}$.

	In the following, we refer to 
	\begin{align*}
		H = H_{0,\mbf k} + H_{\rm dr,\mbf k}
	\end{align*}
	as the light-driven system and to
	\begin{align*}
		H_{\rm st} = H_{0,\mbf k} + H_{\rm ef,\mbf k}
	\end{align*}
	as the static effective model.

	In addition to these unitary contributions we model dissipative processes via Lindblad operators. We introduce three Lindblad operators in the basis that diagonalizes the instantaneous Hamiltonian. The first operator describes decay from the upper to the lower band, with rate $\gamma_{1}=1/T_{1}$. The second operator describes bare dephasing between the upper and the lower band, described by a rate $\gamma_{z}$. We introduce the combined decay rate $\gamma_2$, which descibes the effective dephasing rate, via
	\begin{align}
		\gamma_2=1/T_2=\gamma_1/2+2\gamma_z \label{eq:dephasing}
	\end{align}
	The third Lindblad operator represents single-particle exchange with a fermionic bath of temperature $T$ and chemical potential $\mu$, described by a rate $\gamma_{p} = 1/T_{p}$. 

	We solve the master equation numerically, compute the current for each momentum by taking the trace of the current operator $\mbf j_{y, \mbf k}=ev_{\rm F} \vb{\tau_z {\sigma_x} \mbf e_x + {\sigma_y} \mbf e_y}$ with the density matrix. The contribution to the conductivity of the momentum $\mbf k$ is obtained from the limit $\sigma_{\rm xy}(\mbf k)=\lim_{E_{L} \rightarrow 0} \mbf j_{y, \mbf k}/E_{\rm L}$. We define the contribution to the conductivity dichroism of the momenta $\mbf k$ and $-\mbf k$, via
	\begin{align*}		
		\tilde \sigma_{\rm xy}(\mbf k)&=\frac 1{2A}\, \vb{\sigma_{xy}(\mbf k) +\sigma_{xy}(-\mbf k)}
	\end{align*}
	where $A$ is the lattice size and the full conductivity
	\begin{align*}
		\sigma_{\rm xy}&= \sum_{\mbf k} \tilde \sigma_{\rm xy}\dt
	\end{align*}
	The circular dichroism of the Hall conductivity describes difference of the response for right- and left-handed circularly-polarized light.

	\begin{figure*}[tb]
		\flushleft
		\includegraphics[width=498pt]{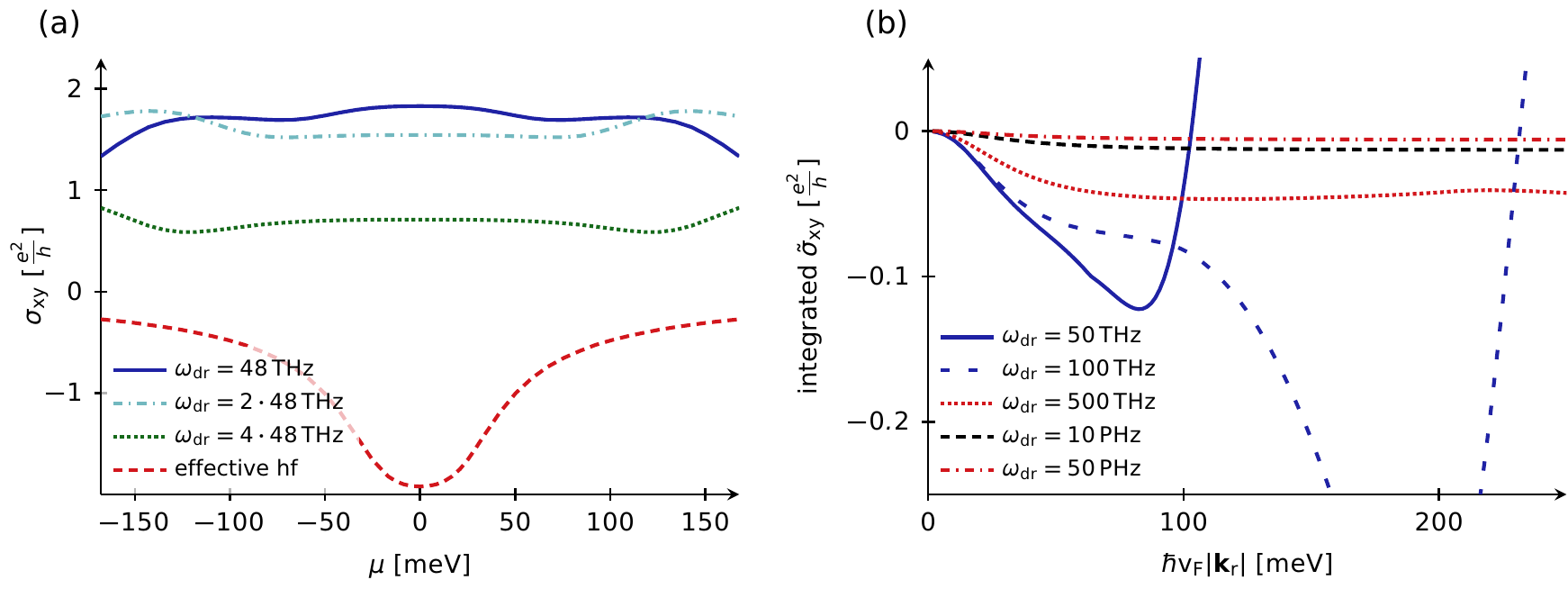}
		\caption{Increasing the driving frequency, while adjusting the driving field $E_{\rm dr}$ to keep the energy gap $\Delta_{\rm hf}=\vb{\hbar v_{\rm F} e E_{\rm dr}}^2/{(\hbar \omega_{\rm dr})^{3}}$ fixed. In both panels we show the circular dichroism of the Hall conductivity. (a) Total conductivity as a function of the chemical potential. For comparison we show the corresponding conductivity of the static effective model, as defined by Eq.~\ref{eq:effHighFreqHamiltonian}. 
		We observe that the Hall conductivity has the opposite sign for low- and high-frequency driving.
		(b) We show the conductivity density $\tilde \sigma_{\rm xy}(\mbf k)$ integrated over all momenta smaller than the threshold value $\abs{\mbf k_{\rm r}}$.
		We find that the contribution of the Dirac point becomes vanishingly small instead of converging towards $-2e^2/h$ when increasing the driving frequency by orders of magnitude.
		The driving frequencies for both panels are indicated in the legends. The electric field strength is $E_{\rm dr}=\usi{20}{\mega\volt\per\meter}$ for $\omega_{\rm dr}=2\pi\cdot\usi{48}{\tera\hertz}$ and is adjusted such that $\Delta_{\rm hf}$ has the same value for all driving frequencies $\omega_{\rm dr}$. The remaining parameters for all panels are $T_1=\usi{100}{\femto\second}$, $T_2=\usi{20}{\femto\second}$, $T_p=\usi{25}{\femto\second}$, $T=\usi{80}{\kelvin}$ and $E_{\rm L}=\usi{840}{\volt\per\meter}$. All observables are shown after a steady state is achieved for a tanh-type ramp of the driving field strengt, for panel (a) $t_{r,\rm dr}=\usi{1}{\pico\second}$ and $\sigma_{r,\rm dr}=\usi{0.5}{\pico\second}$, while for panel (b) $t_{r,\rm dr}=\usi{0.5}{\pico\second}$ and $\sigma_{r,\rm dr}=\usi{0.25}{\pico\second}$.}
		\label{fig:highFreqLimit}
	\end{figure*}

	As a central tool to investigate the Hall conductivity of the Dirac point and individual resonances we show the conductivity dichroism $\tilde\sigma_{\rm xy}(\mbf k)$ in Fig.~\ref{fig:momentumResolvedConductivity}(a) and its radially integrated value in Fig.~\ref{fig:momentumResolvedConductivity}(b), i.e. $\int_0^{\abs{\mbf k_{\rm r}}} \tilde\sigma_{\rm xy}(\mbf k)$. In addition to the contribution of the Dirac point there are contributions whenever the driving frequency is resonant with the band gap, i.e.~$\omega_{\rm dr}=2v_Fk$. For small field strength $E_{\rm dr}$, the contributions to the Hall conductivity dichroism are localised around each of these resonances, depicted by dashed lines in Fig.~\ref{fig:momentumResolvedConductivity}(b). The initial decrease in Fig.~\ref{fig:momentumResolvedConductivity}(b) corresponds to the contribution of the Dirac point. It is much smaller than $-2e^2/h$, which is the result of the static effective model. As an estimate for the contribution of the Dirac point we can define
	\begin{align}
		\sigma_{\rm Dirac}&=\int_0^{k=\omega_{\rm dr}/ (4v_{\rm F})} \D^2 \mbf k \; \tilde \sigma_{xy} \label{eq:diracPointContributionHallCond} \com
	\end{align}
	which we use below.

\section{High-frequency limit}\label{highFrequencyLimit}
	As a first approach we consider increasing the driving frequency, and compare it to the quantized integer conductivity of the Floquet Hamiltonian intepreted as a static Hamiltonian with the gap term in Eq.~\ref{eq:effHighFreqHamiltonian}. Changing the driving frequency only, implies a reduction of the gap at the Dirac point, which scales as $\Delta_{\rm hf}\propto E_{\rm dr}^2/\omega_{\rm dr}^3$, see Eq.~\ref{eq:effHighFreqHamiltonian}. Hence, for increasing driving frequencies the gap at the Dirac point is reduced. We therefore consider the case of increasing the driving field strength $E_{\rm dr}$ such that the value of $\Delta_{\rm hf}$ remains constant. We display this scenario in Fig.~\ref{fig:highFreqLimit}(a), where we show the Hall conductivity dichroism as a function of the chemical potential $\mu$ for several different driving frequencies $\omega_{\rm dr}$. We compare the conductivity of the light-driven system with the static effective high-frequency limit which we model by using the effective static Hamiltonian from Eq.~\ref{eq:effHighFreqHamiltonian}. We find that the conductivity of the light-driven system has the opposite sign of the static effective system. Additionally we observe in Fig.~\ref{fig:highFreqLimit}(a) that for the given range of frequencies the conductivity of the light-driven system decreases slowly for increasing driving frequency.

	Given that higher driving frequencies and larger driving field strengths are numerically challenging, we focus on the contribution of the Dirac point in Fig.~\ref{fig:highFreqLimit}(b).
	For $\omega_{\rm dr}=2\pi\cdot\usi{50}{\tera\hertz}$ and $\omega_{\rm dr}=2\pi\cdot\usi{100}{\tera\hertz}$ we see the onset of the first resonance, at $\hbar v_{\rm F} \abs{\mbf k_{\rm r}}\approx \usi{100}{\milli\eV}$ and $\hbar v_{\rm F} \abs{\mbf k_{\rm r}}\approx \usi{200}{\milli\eV}$, respectively. For larger driving frequencies the plateau at larger threshold momenta gives the Dirac-point contribution to the conductivity. In the static high-frequency limit it is expected to converge towards $-2e^2/h$. However, the Hall conductivity of the light-driven system is notably different than the hypothesized limit of equating the Floquet states to energy states of a static Hamiltonian. We observe in Fig.~\ref{fig:highFreqLimit}(b) for increasing driving frequencies that the conductivity reduces to 0 instead. This result is consistent with the formation of a high-temperature state with almost equal occupation of the lower and upper graphene band. In Ref.~\cite{nuske_graphene} we have seen that the Hall conductivity can be approximated by weighting Berry curvature and band velocity with the corresponding band occupations. Since both bands have opposite Berry curvature and band velocity the resulting Hall conductivity vanishes.	

\section{Geometric Hall conductivity for low dissipation}\label{lowDissipation}
	Next we consider reducing dissipation. We find that for fixed driving frequency the dissipation greatly inhibits the Hall-conductivity contribution of the Dirac-point gap. The contributions of resonances are enhanced by some dissipation mechanisms, see App.~\ref{app:reducingDephasing}. We can identify a regime of low dissipation and intermediate electric field strengths where the Hall-conductivity contribution of the Dirac point approaches $-2e^2/h$.

	In this and the following section we switch off the coupling to the back gate ($T_{\rm p}=\infty$). Hence, we work at fixed particle number, enforcing unit occupation of each momentum mode. The computations with decoupled back gate are more efficient and hence allow to investigate a wider parameter regime. Apart from being numerically more feasible we have several other reasons motivating the decoupled back gate: experimentally this could be achieved by using free-standing graphene, i.e.~graphene without a substrate. Also we confirm in App.~\ref{app:finiteBackgate} that at zero chemical potential the results are qualitatively similar to those with coupled back gate. Finally, it is an interesting question on its own to ask whether decay-type damping inhibits the geometric Hall conductivity of the gap at the Dirac point. In addition to decoupling the back gate we consider only cases where $T_2=2T_1$, which corresponds to small dephasing-type dissipation. We explain in App.~\ref{app:reducingDephasing} that this is beneficial for obtaining a large contribution of the Dirac point to the Hall conductivity.

	\begin{figure}[tb]
		\flushleft
		\includegraphics[width=246pt]{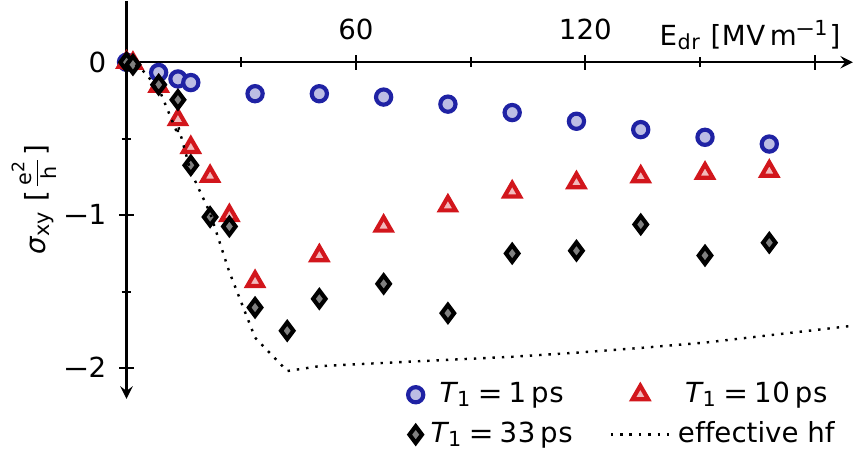}
		\caption{Contribution of the Dirac point to the Hall conductivity dichroism, as defined in Eq.~\ref{eq:diracPointContributionHallCond}, for different damping time scales as indicated in the legend. For comparison the dotted line shows the static effective model, as defined by Eq.~\ref{eq:effHighFreqHamiltonian}, at $T_1=\usi{1}{\pico\second}$. Note that the static effective model has no strong dependence on dissipation, for driving field strengths that are sufficiently large so that the induced energy gap is larger than the dissipative broadening and the temperature. We observe a regime of small dissipation and intermediate driving field strengths where the Hall conductivity of the light-driven system approaches the effective static result of $-2e^2/h$. We use $\omega_{\rm dr}=\usi{2\pi\cdot 200}{\tera\hertz}$, $T_2=2T_1$, $T_p=\infty$, $T=\usi{80}{\milli\kelvin}$, $E_{\rm L}=\usi{1.7}{\kilo\volt\per\meter}$ and $\mu=0$. All observables are shown after a steady state is achieved following a tanh-type ramp-up function of the driving field strength with time scales  $t_{r,\rm dr}=\usi{1}{\pico\second}$ and $\sigma_{r,\rm dr}=\usi{0.5}{\pico\second}$. The simulations for the low dissipation cases are numerically challenging which is the origin of the numerical noise in the $T_1=\usi{33}{\pico\second}$ data.}
		\label{fig:lowDrivRecoverHaldane}
	\end{figure}

	We show the driving-field-strength dependence of the Dirac-point contribution for different $T_1$ in Fig.~\ref{fig:lowDrivRecoverHaldane}. In the effective high-frequency limit the contribution of the Dirac point to the Hall conductivity is small for low electric field strength, increases to approximately $-2e^2/h$ and then reduces again for even larger field strengths. At low field strength the size of he gap is small compared to dissipation and temperature resulting in a partial cancellation of the conductivity. The origin of the reduction for large field strength is that the conductivity contribution spreads out in momentum space and hence extends beyond the regime that we associate with the Dirac point. At intermediate field strengths we find a plateau near $-2e^2/h$ for the effective high-frequency system.
	
	For the light-driven system we see in Fig.~\ref{fig:lowDrivRecoverHaldane} that decay-type damping has a strong influence on the Hall conductivity of the Dirac point. 
	The Dirac-point contribution is reduced significantly for large decay-type damping, i.e.~small $T_1$. For larger values of $T_1$ the contribution of the Dirac point agrees with the one computed for the effective high-frequency system for small values of the driving field strength. When further increasing $T_1$ the range in which the two contributions agree extends to larger and larger field strength.
	For $T_1=\usi{33}{\pico\second}$ we nearly recover the quantized value of $-2e^2/h$ at intermediate field strengths. We expect that the Hall-conductivity contribution of the Dirac point converges towards its high-frequency limit when reducing dissipation. 	

\section{Geometric Hall conductivity in the transient response}\label{transientResponse}
	\begin{figure}[tb]
		\centering
		\includegraphics[width=246pt]{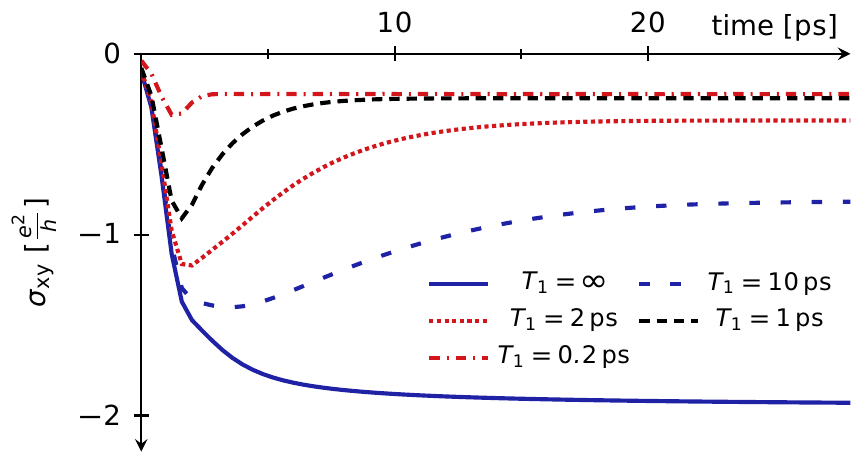}
		\caption{Contribution of the Dirac point to the Hall conductivity dichroism, as defined in Eq.~\ref{eq:diracPointContributionHallCond}, as a function of time $t$ for several dissipation times $T_1$. We use $E_{\rm dr}=\usi{84}{\mega\volt\per\meter}$, $\omega_{\rm dr}=\usi{2\pi\cdot 200}{\tera\hertz}$, $T_2=2T_1$, $T_p=\infty$, $T=\usi{80}{\milli\kelvin}$, $E_{\rm L}=\usi{1.7}{\kilo\volt\per\meter}$, $\mu=0$ and the ramp-up time scales $t_{r,\rm dr}=\usi{1}{\pico\second}$ and $\sigma_{r,\rm dr}=\usi{0.5}{\pico\second}$.
		 } 
		\label{fig:transientHallConduct}
	\end{figure}
	As a third comparison, we consider the contribution of the Dirac point to the Hall conductivity during the transient response before the system equilibrates. We show the time-resolved contribution of the Dirac point to the Hall conductivity in Fig.~\ref{fig:transientHallConduct}. The net Hall conductivity is negative for all dissipation strengths and times. 
	For vanishing dissipation, corresponding to $T_1 = \infty$, the Hall conductivity approaches $-2e^2/h$. 
	For non-zero, but small dissipation we find that the Hall conductivity approaches $-2e^2/h$ before it equilibrates at a smaller value. The dissipative process results in a population of both the upper and the lower band at the Dirac point, which gives a partial cancellation of the Hall conductivy. For stronger dissipation, i.e. smaller $T_1$, both the transient and the equilibrium Hall conductivity is suppressed as compared to weaker dissipation. 

\section{Summary and Outlook}\label{summary}
	We have investigated the Hall conductivity of light-driven graphene for a wide range of driving frequencies and dissipation strengths. As a central focus, we have compared the Hall conductivity of the light-driven system to a hypothetical limit that we refer to as the static effective model, in which the Floquet quasi-energies are interpreted as energies of a static Hamiltonian, to determine if this static effective limit provides a valid approximation for the light-driven, dynamical system. We have focused on the contribution of the Dirac point on the Hall conductivity which is expected to show a quantized response of $-2e^2/h$ for large driving frequencies and vanishing dissipation for the static effective model. We find that the steady-state contribution of the Dirac point is crucially influenced by dissipation. As a central focus, we have compared the Hall conductivity of the light-driven system to a hypothetical limit that we refer to as the static effective model, in which the Floquet quasi-energies are interpreted as energies of a static Hamiltonian, to determine if this static effective limit provides a valid approximation for the light-driven, dynamical system. For finite dissipation we find that the Hall conductivity vanishes in the high-frequency-driving limit. Instead we identify a regime of small dissipation and intermediate driving frequencies and field strengths where we recover the value of $-2e^2/h$ approximately. 

	Our work paves the way for a more complete understanding of periodically driven solids in the dissipative regime. We emphasize that similar conclusions as we present here will apply to light-driven materials in general. This implies that, in general, a prediction for a Floquet-enegineered material cannot by achieved by determining the Floquet states and using the  static effective approximation. Rather, the inclusion of dissipative processes creates a steady state, for which the magnitude of the observables depend on the type and magnitude of the dissipative processes. We identify a narrow regime, in which the static effective model gives an acceptable approximation of the steady state, and point out that the transient state of the light driven material is approximated by the static effective model, for sufficiently small dissipation. With these insights, we refine and advance Floquet engineering in general.

 \acknowledgments{
We thank Lukas Broers, James McIver and Gregor Jotzu for fruitful discussions.
We acknowledge support from the Deutsche Forschungsgemeinschaft through the SFB 925. This work is supported by the Cluster of Excellence 'CUI: Advanced Imaging of Matter' of the Deutsche Forschungsgemeinschaft (DFG) - EXC 2056 - project ID 390715994. M.N.~acknowledges support from Stiftung der Deutschen Wirtschaft. 
}

\begin{appendix}
	\section{Reducing the dephasing-type dissipation}\label{app:reducingDephasing}
		\begin{figure}[tb]
			\centering
			\includegraphics[width=246pt]{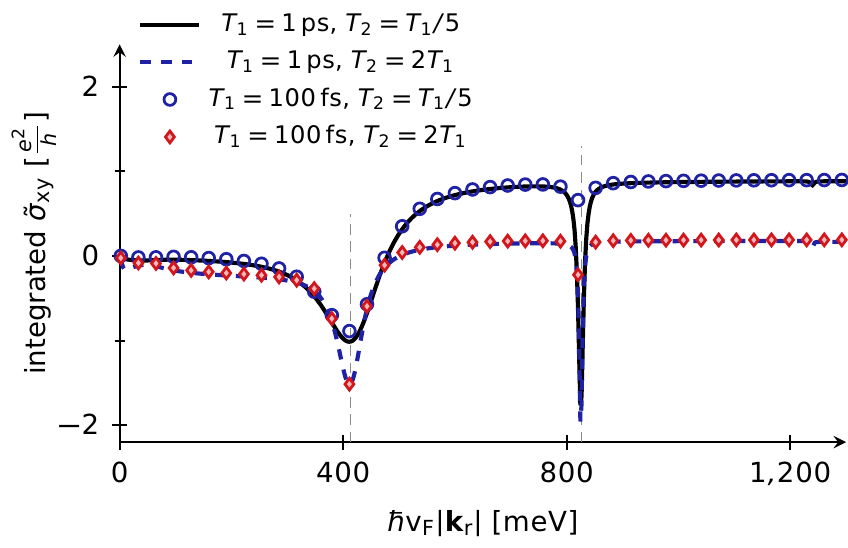}
			\caption{Circular dichroism of the integrated Hall conductivity densities for several different values of the dissipation as indicated in the legend. We show the conductivity density $\tilde \sigma_{\rm xy}$ integrated over all momenta smaller than the threshold value $\abs{\mbf k_{\rm r}}$. The first and second resonance are indicated by dashed lines. We use $E_{\rm dr}=\usi{84}{\mega\volt\per\meter}$, $\omega_{\rm dr}=\usi{2\pi\cdot 200}{\tera\hertz}$, $T_p=\infty$, $T=\usi{80}{\milli\kelvin}$, $E_{\rm L}=\usi{1.7}{\kilo\volt\per\meter}$ and $\mu=0$. All observables are shown after a steady state is achieved for a tanh-type ramp of the driving field strength with $t_{r,\rm dr}=\usi{1}{\pico\second}$ and $\sigma_{r,\rm dr}=\usi{0.5}{\pico\second}$.}
			\label{fig:conductivityKResDifferenDamping}
		\end{figure}

		Here we discuss the role of dephasing-type damping, i.e.~$T_2$. We observe in Fig.~\ref{fig:conductivityKResDifferenDamping} that the relative scale of $T_1$ and $T_2$ crucially influences the Hall-conductivity contributions of the Dirac point and the first resonance. Large dephasing type damping, i.e.~small $T_2$, suppresses the contribution from the Dirac point, while at the same time enhancing the contribution of the first resonance. Compared to this effect the overall time scale of damping at fixed $T_2/T_1$ plays a negligible role for $\usi{100}{\femto\second}<T_1<\usi{1}{\pico\second}$. We note that there is an upper bound $T_2/T_1\leq 2$, since there is a finite amount of dephasing for each decay process. This can also be motivated from Eq.~\ref{eq:dephasing} in the main text, where we see, when setting the bare dephasing-type damping $\gamma_z=0$, that at low temperature
		\begin{align*}
			\frac{1}{T_2}&=\Gamma=\frac{\gamma_1}{2}\approx \frac{1}{2T_1}\dt
		\end{align*}
		Our results motivate the use of the optimal case of $T_2=2T_1$ in the main text.

	\section{Time resolved occupations}
		\begin{figure*}[h!tb]
			\centering
			\includegraphics[width=404pt]{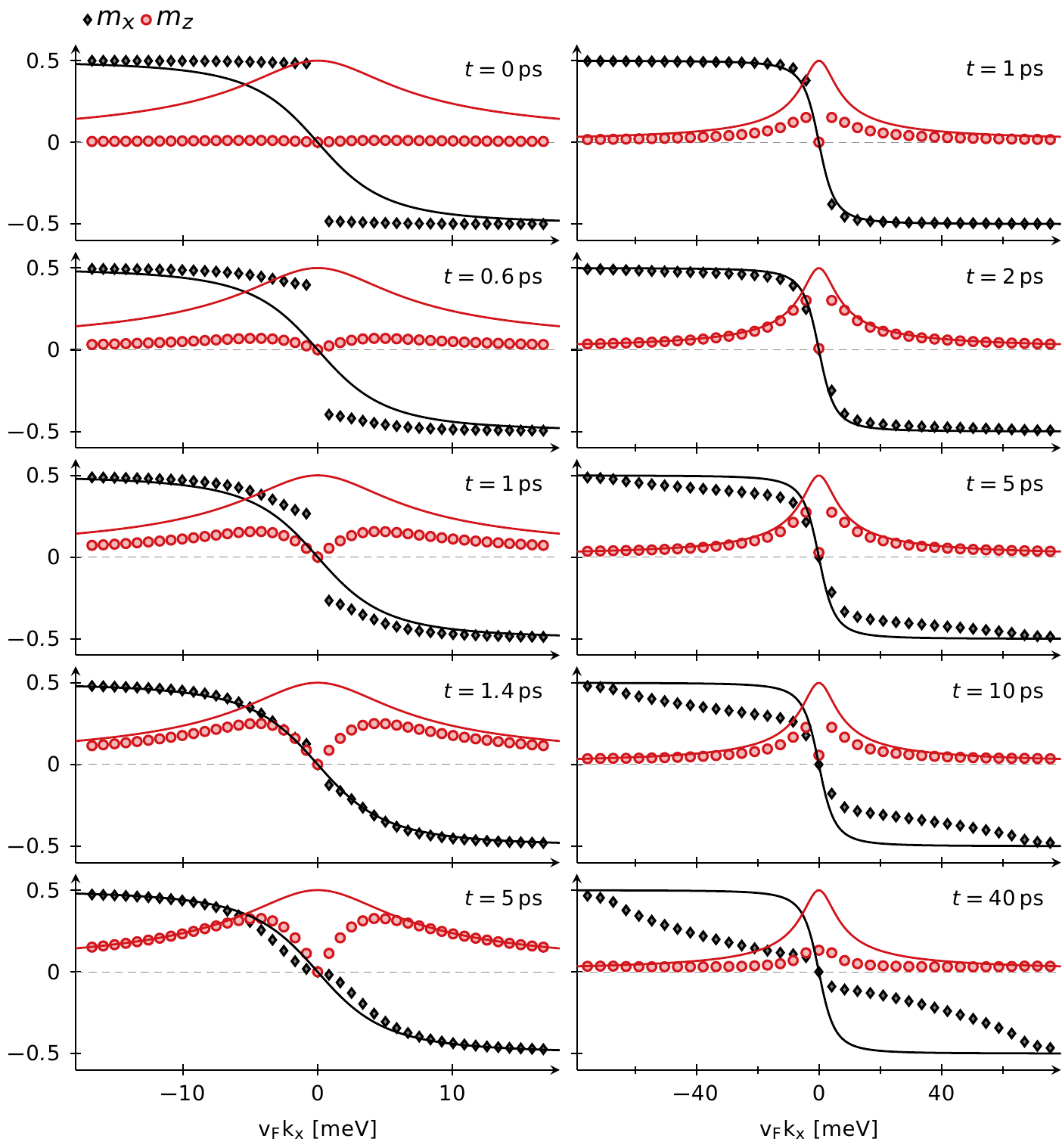}
			\caption{Coefficients of the density matrix for a cut along the $k_x$-direction at different snapshots in time as indicated in the legend. For comparison we show the ground state of the effective high-frequency system with solid lines. We show $m_x$ by black triangles and $m_z$ by red circles, we do not show $m_y$ since it is negligible along the $k_x$-direction. The left column shows the undamped system while the right column shows $T_1=\usi{10}{\pico\second}$, $T_2=2T_1$ and $T_p=\infty$. For all panels $E_{\rm dr}=\usi{84}{\mega\volt\per\meter}$, $\omega_{\rm dr}=2\pi\cdot\usi{200}{\tera\hertz}$, $T=\usi{80}{\milli\kelvin}$, $E_{\rm L}=0$ and the ramp-up time scales are $t_{r,\rm dr}=\usi{0.5}{\pico\second}$ and $\sigma_{r,\rm dr}=\usi{0.25}{\pico\second}$.}
			\label{fig:mxmzOfTime}
		\end{figure*}

		In order to get an intuitive understanding of the transient regime we first consider the time evolution of the density matrix without an applied longitudinal field. When decoupling the back gate ($T_p=\infty$) the system has unit filling and hence can be described by a $2$\hspace{0.5pt}$\times$\hspace{0.5pt}$2$ density matrix
		\begin{align*}
			\rho_\bk &=1/2 + m_x(\bk) \sigma_x + m_y(\bk) \sigma_y +m_z(\bk) \sigma_z \dt
		\end{align*}
		We show the coefficients of the density matrix for a cut along the $k_x$-direction in Fig.~\ref{fig:mxmzOfTime}. We do not show $m_y(\bk)$ since it is negligible for $k_y=0$. We first consider the undamped case which is shown in the left column of Fig.~\ref{fig:mxmzOfTime}. Initially the density matrix correctly shows the dependence for undriven graphene where $m_x=-k_x/\abs \bk$ and $m_z=0$. Subsequently the system evolves towards the ground state of the driven effective high-frequency system shown by solid lines in Fig.~\ref{fig:mxmzOfTime}. For large momenta it reaches this value after about $\usi{5}{\pico\second}$. Here the quench is adiabatic since the time scale of the quench of about $t_{\rm quench}=\usi{1}{\pico\second}$ ($\hbar/t_{\rm quench}\approx \usi{0.7}{\milli\eV}$) is slow compared to the bandwidth of undriven graphene $2\epsilon_\bk=2\hbar v_F k$. Directly at the Dirac point the bands of undriven graphene touch and hence the quench can not be adiabatic. The time scale of the quench determines the range around the Dirac point where the quench is nonadiabatic. For longer quench time scales the region where the response deviates from the ground state of the high-frequency system is more confined around the Dirac point. Finite $m_z$-component around the Dirac point is important for having finite Berry curvature and hence a finite Hall conductivity. 
		With dissipation momentum points that are nonadiabatic may relax towards the ground-state value of the high-frequency system, see right column of Fig.~\ref{fig:mxmzOfTime}. At the same time dissipation leads to a reduction of all coefficients of the density matrix. We find that this reduction is stronger for larger dissipation. This overall reduction also reduces the Hall conductivity. In the right column of Fig.~\ref{fig:mxmzOfTime} we consider a regime where the quench duration is faster than the relaxation time scale $T_1$. Then there are two different time scales. First the system approaches the ground state of the high-frequency system for all momentum modes that are adiabatic within the first $\usi{5}{\pico\second}$. Then on a longer time scale the density matrix components of all momenta approach the steady state which is reached after about $\usi{40}{\pico\second}$.
		At intermediate times the $m_z$ component of the density matrix at the Dirac point has already nonzero value while at the same time the density-matrix components of larger momenta are still close to the effective high-frequency system. This is the regime where we expect the quantized Hall response of the Dirac point.	

	\section{Comparison of the {$2$\hspace{0.5pt}$\times$\hspace{0.5pt}$2$} and {$4$\hspace{0.5pt}$\times$\hspace{0.5pt}$4$} model} \label{app:finiteBackgate}
		\begin{figure}[tb]
			\centering
			\includegraphics[width=\linewidth]{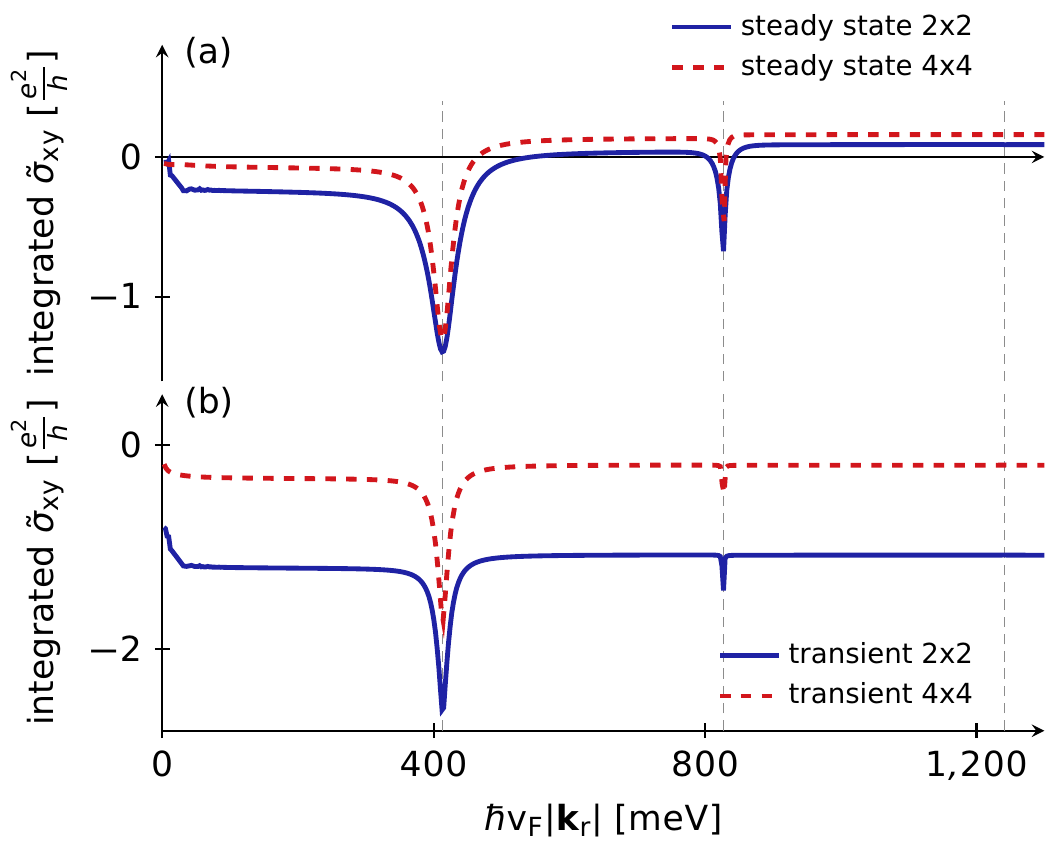}
			\caption{Comparison of the integrated Hall conductivity dichroism for the $2$\hspace{0.5pt}$\times$\hspace{0.5pt}$2$ and the $4$\hspace{0.5pt}$\times$\hspace{0.5pt}$4$ model for low dissipation. We show the conductivity density $\tilde \sigma_{\rm xy}$ integrated over all momenta smaller than the threshold value $\abs{\mbf k_{\rm r}}$. Panel (a) shows the conductivity after a steady state has been achieved for a tanh-type ramp of the driving field strength, while panel (b) shows the transient response. We use $E_{\rm dr}=\usi{30}{\mega\volt\per\meter}$, $\omega_{\rm dr}=\usi{2\pi\cdot 200}{\tera\hertz}$, $T_1=\usi{10}{\pico\second}$, $T_2=\usi{2}{\pico\second}$, $T_p=\usi{4}{\pico\second}$, $T=\usi{80}{\milli\kelvin}$, $E_{\rm L}=\usi{1.7}{\kilo\volt\per\meter}$, $\mu=0$, $t_{r,\rm dr}=\usi{0.5}{\pico\second}$ and $\sigma_{r,\rm dr}=\usi{0.25}{\pico\second}$. The first, second and third resonance are indicated by dashed lines. }
			\label{fig:compare2244WeakDiss}
		\end{figure}
		As we have noted, Secs.~\ref{lowDissipation} and \ref{transientResponse} in the main text have used the effective $2$\hspace{0.5pt}$\times$\hspace{0.5pt}$2$-system, i.e.~have considered the system with fixed particle number $T_p\To\infty$. Here we compare these results to the full $4$\hspace{0.5pt}$\times$\hspace{0.5pt}$4$ system, see Fig.~\ref{fig:compare2244WeakDiss}. We see that finite back gate damping $T_p$ within the $4$\hspace{0.5pt}$\times$\hspace{0.5pt}$4$ model significantly reduces the contribution of the Dirac point. While the overall magnitude is reduced the conductivity in the transient regime is still larger than the corresponding steady-state result.	

	\end{appendix}

\bibliography{bib_paper}

\end{document}